\documentclass[twocolumn,10pt]{article}

\usepackage{hyperref}

\usepackage{graphicx}
\usepackage{booktabs}

\usepackage[disable]{todonotes}

\usepackage{enumitem}  

\usepackage{float}  

\title{MMS Player: an open source software for parametric data-driven animation of Sign Language avatars}

\author{Fabrizio Nunnari, Shailesh Mishra, Patrick Gebhard \\
fabrizio.nunnari@dfki.de, mshailesh2018@gmail.com, patrick.gebhard@dfki.de \\
 German Research Center for Artificial Intelligence (DFKI)}	

\date{}

\begin{document}

\maketitle

\begin{abstract}
This paper describes the MMS-Player, an open source software able to synthesise sign language animations from a novel sign language representation format called MMS (MultiModal Signstream). The MMS enhances gloss-based representations by adding information on parallel execution of signs, timing, and inflections. The implementation consists of Python scripts for the popular Blender 3D authoring tool and can be invoked via command line or HTTP API. Animations can be rendered as videos or exported in other popular 3D animation exchange formats.
The software is freely available under GPL-3.0 license at \url{https://github.com/DFKI-SignLanguage/MMS-Player}.
\end{abstract}


\subsection*{Keywords}
sign language, synthesis, avatar animation, inflection, MMS, multi-modal signstream.

\begin{figure*}
  \includegraphics[width=0.48\textwidth]{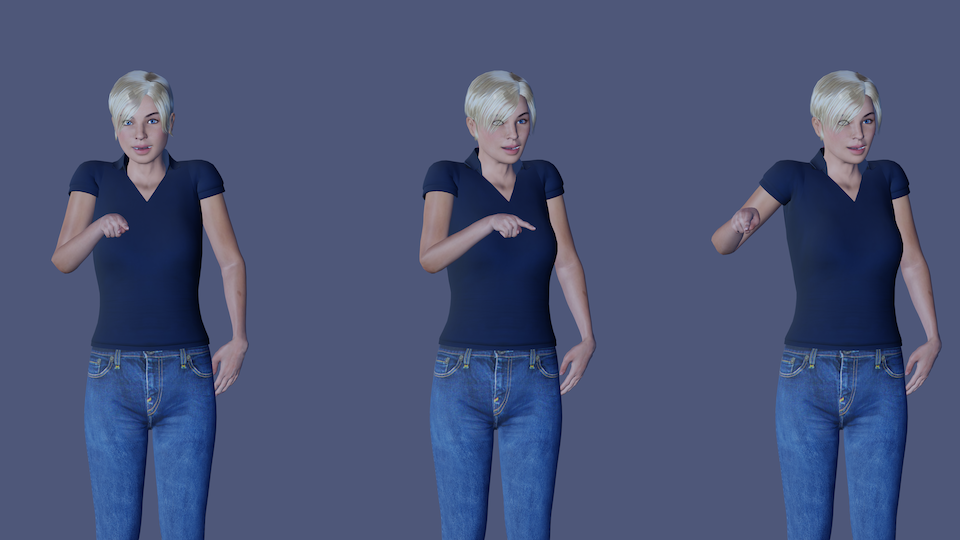}
  \includegraphics[width=0.48\textwidth]{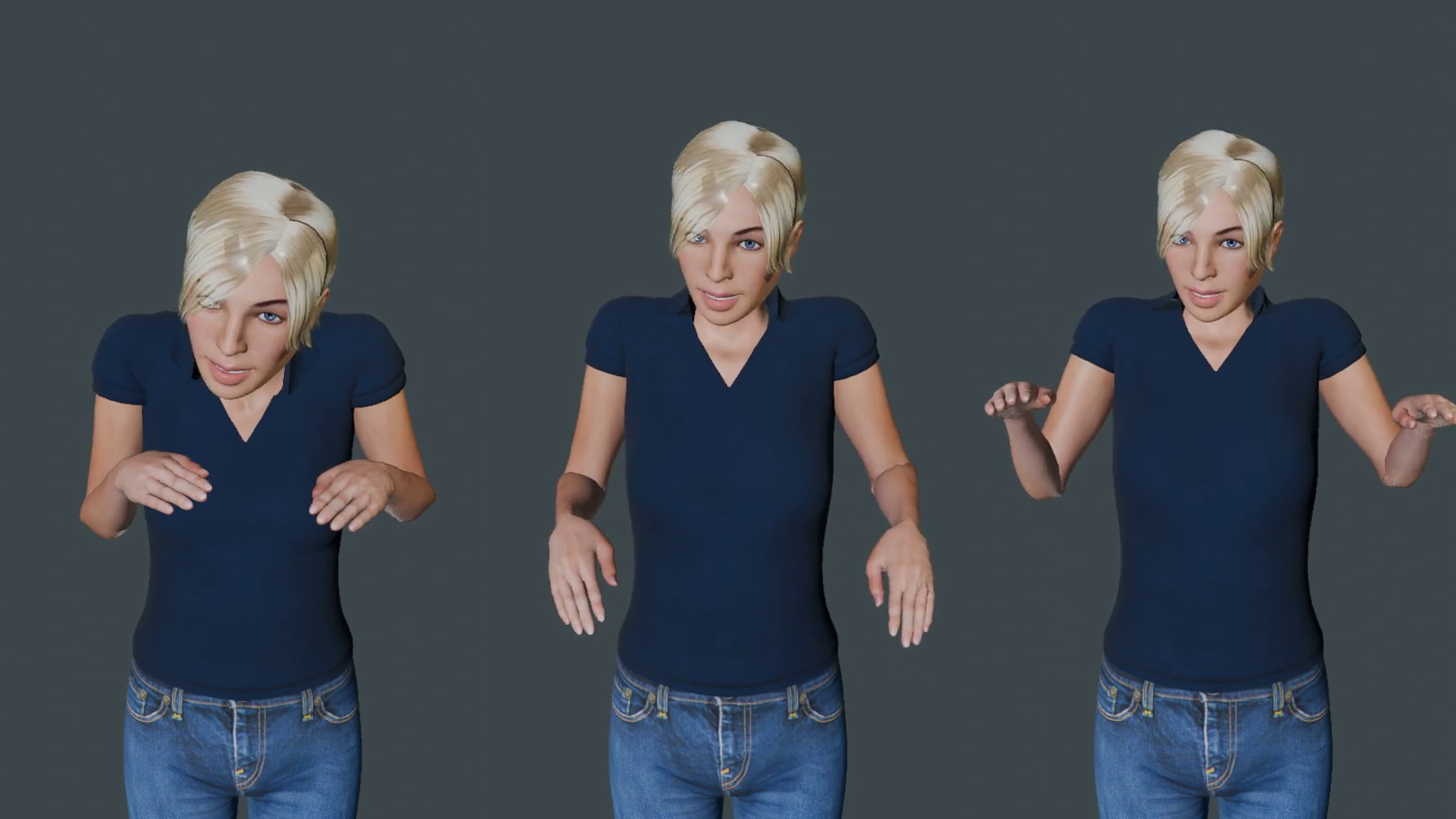}
  \caption{Thanks to its inflection capabilities, the MMS-Player can realize new contextualized signs from existing ones. Left: the sign INDEX in its citation form; rotating torso, arm, and hand independently to indicate something on the left side; and orienting the torso to the left but pointing to the right. Right: the sign NICHT (\emph{not} in German sign language) performed smaller (as ``whispering'' to someone close); in its citation form; and bigger.}
  \label{fig:teaser}
\end{figure*}


\maketitle

\section{Introduction}


Sign language (SL) is the native communication language of about 70 million people all over the world, spread over more than 300 different sign languages \cite{united-nations_international_2024}.
Despite 20+ years of research in the field, the automatic translation between spoken/written to signed languages is still at early stage (e.g., see the results of the latest MT challenges \cite{avramidis_linguistically_2022,avramidis_challenging_2023}).


In this paper/demo, we focus on SL \emph{synthesis} (text to animation) using procedurally animated avatars (3D virtual characters).

In particular, we present two main contributions.
First (section~\ref{sec:mms}), we describe the current version of the MMS (MultiModal Signstream): a SL machine-readable representation that augments the concept of gloss sequence with extra information on sign parallelization, timing, and motion inflections.
Second (section~\ref{sec:sw-architecture}), we present the software architecture of a fully working version of an ``MMS realizer'', i.e., a software capable of taking an MMS instance as input and producing SL animations based on a 3D virtual character.
The system has been also preliminarily evaluated by a group of native deaf and interpreters (section~\ref{sec:evaluation}).\\

The definition of MMS comes from the need to express sign language in a machine-readable format that can be the target of text-to-MMS (or vice versa) automated translation systems. Most of the existing work relies on ``gloss sequence'', i.e., a sequence of words whose meaning is close to the observed corresponding sign. However, glosses are not enough. SLs manifest complex phenomena like the parallel execution of multiple signs, modulation of the execution speed, pauses, and ``inflections'', i.e., adaptation of the execution of the sign to the context of the expressed message. For example, a sign can be expanded or shrunk to express the adjectives \emph{big} and \emph{small}; signs must be \emph{relocated} in the signing space; the torso needs to rotate and lean to realize \emph{role taking}.

To accommodate the need of ``parameterization'' of signs, some systems rely on SL representations like HamNoSys \cite{hanke_hamnosys-representing_2004} or SignWriting \cite{sutton_lessons_1995}, but realizing SL directly from those descriptors leads to robotic and unrealistic movements.
In contrast, more natural motion can be obtained by recording the motion of humans while performing signs.
However, historically, systems based on MoCap (Motion Capture) animation data were considered non-parameterizable, and only able to concatenate signs as they were recorded (\emph{citation form}).

The MMS-Player challenges this conviction and proposes a system capable to perform ``motion editing'' of recorded signs and modify them on-the-fly to realize the many dynamic phenomena of SL.
As an example, Figure \ref{fig:teaser} shows the execution of the sign INDEX (a deictic multi-purpose gesture) and NICHT (in German Sign Language, DGS) in their citation form and with two inflected versions.

\begin{figure*}
    \centering
    \includegraphics[width=1.0\textwidth]{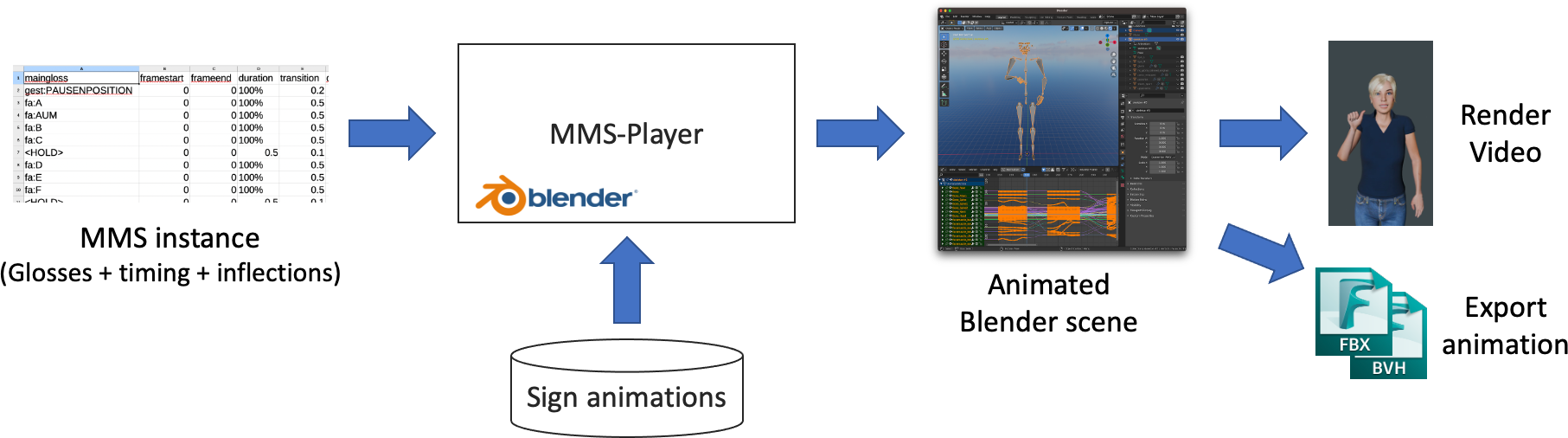}
    \caption{The MMS-Player conceptual realization pipeline.}
    \label{fig:pipeline}
\end{figure*}

Figure \ref{fig:pipeline} shows the conceptual realization pipeline of the MMS-Player.
Given an MMS instance, the MMS-Player reads one-by-one all its rows, each containing a gloss-ID, timing information, and inflection information. For each gloss, the corresponding animated sign is read from a dictionary, adjusted in time/duration, and ``inflected''. Each inflected gloss is then serialized into a timeline, which can be either rendered as video or exported in many 3D animation formats.

Hence, the MMS-Player is meant to be the second stage of a text-to-MMS translation pipeline.
The first stage is currently still under investigation, and thus we cannot yet measure the quality of a translation from text to MMS, but we see at least two key advantages in breaking a full translation pipeline in two stages. First, translating text-to-MMS is supposedly an easier task than translating directly into high-density motion stream (i.e., requires fewer data points). Second, we believe that, through a dedicated 3D GUI, MMS instances could be manually created and corrected by humans. The long-term goal is to develop a dedicated CAT-tool (computer-aided translation), integrating the MMS-Player, where professional translators could correct and preview automatically generated MMS instances. In an iterative loop, new human-verified text-to-MMS data points could then be collected and used for training improved translation models.\\


This work is the culmination in a stable and functioning public implementation of the concept already introduced by Nunnari et al. in 2023 \cite{nunnari23SLTAT-InflectionParameters}.
With respect to the initial proposal, in this version, in addition to the hands, the full body inflection (shoulder, torso, and head) is also implemented together with the timing routines. We thus present the details of the configurations and the algorithms conceived for the realization. The software is now available for download in an open source repository\footnote{\url{https://github.com/DFKI-SignLanguage/MMS-Player}} and a demonstration video is available online\footnote{\url{https://www.youtube.com/watch?v=MxiFsxXukZw}}.
Finally, we report the results of a preliminary user study that helped us identifying priorities and guidelines for future development.

\section{Related work}




In their 2020 review, Naert et al.~\cite{naert_survey_2020} claim that ``the signs in their citation form are not enough for utterance synthesis: a simple concatenation of those signs to create an utterance does not do justice to the richness of SL and can generate incorrect utterances''. In fact, they propose, too, a system for motion editing of recorded signs through the combination of motions from different body parts.
Unlike their approach, the MMS-Player is able not only to merge different animations, but also implements a set of functions to programmatically edit existing motion.

Recently, translation to sign language has been addressed using photorealistic video synthesis. One of the first proposals, by Stoll et al.~\cite{stoll_sign_2018}, first translates text into full-body animation data sequences, and then renders a video via generative adversarial networks (GANs). In so doing, 2D body landmarks animation data is in fact the intermediate representation of sign language.
A follow-up of that proposal has evolved into a commercial solution: SignStream\footnote{\url{https://www.signapse.ai/signstream-api}} by the firm Sygnapse.

Despite the promising results, GAN-based approaches still manifest several disadvantages with respect to 3D avatars.
First, they suffer from visual artifacts (such as hallucinations in the movement of the fingers and sudden changes in brightness) due to their statistical approach and hard-to-manage variance in the recorded training data.
Second, the output of a translation cannot be corrected by a human operator, who should go through a slow process of pixel-level manual editing.
Finally, GAN-based videos cannot be applied in interactive 3D environments, such as movable personal interpreters using emerging augmented reality technologies \cite{nunnari_towards_2023}.

For these reasons, we believe that 3D real-time avatars can still be a valuable alternative to video synthesis.\\

Aziz et al.\ \cite{aziz_evolution_2023} present a comprehensive list of 47 avatar-based sign synthesis systems, with screenshots, covering the last 40 years of research.
Of all the surveyed systems, only a few kept on being updated and very few are actually available to the research or public community.

Among them, VLibras\footnote{\url{https://www.vlibras.gov.br/}} \cite{mgisp_vlibras_2025} is a working system for translations on the Brazil Government web sites. However, based on our inspections, the system does not support inflections, and it is based on the Unity 3D engine, binding developers to proprietary licenses.

One of the oldest and yet still-working software is JASigning\footnote{\url{http://vh.cmp.uea.ac.uk/index.php/JASigning_Demos}} \cite{elliott_towards_2010}. It is based on a javascript implementation and is freely available. However, it was architected for a pure synthesis approach, taking as input HamNoSys descriptors \cite{hanke_hamnosys-representing_2004} and thus producing highly robotic motion.

One of the most popular and advanced SL synthesis systems is the research community is the Paula avatar \cite{mcdonald_automated_2016}.
It is able to realize SL using both synthetic and data-driven animation as input, and motion can be edited to the finest level of detail of body and face (e.g., \cite{wolfe_exploring_2018}).
The system is used in advanced studies jointly with sign language experts to increase the expressivity and correctness of the synthesis, but does not focus on the integration within a full text-to-sl translation pipeline.

To our knowledge, the MMS-Player is the first ever available implementation of a SL synthesis system which is completely open source (based on the GPL-3.0 license) and which support a data-driver generation of sign language utterances.

\section{The MMS structure}
\label{sec:mms}

\begin{table*}
\small
\centering
\begin{tabular}{l|l|p{0.5\textwidth}}
\toprule
MMS column & Category & Description \\
\midrule
maingloss & Glosses & The main gloss-ID, specifying which animation has to be played on the full body and face. \\
domgloss & Glosses & The gloss-ID of the animation that replaces the main gloss animation on the dominant arm. \\
ndomgloss & Glosses & Same but for the non-dominant arm. \\
framestart & Timing & The starting time of the gloss playback on the timeline.\\
frameend & Timing & The ending time of the gloss playback on the timeline.\\
transition & Timing (Alternative) & The delay from the previous gloss.\\
duration & Timing (Alternative) & The duration of this gloss of the timeline.\\
{[} n]domhandreloc[ as][xyz] & Inflection (Hands trajectory) & Inflection of the (non)-dominant hand trajectory in terms of translation, rotation, scaling. \\
{[} n]domhandrot[xyz] & Inflection (Hands rotation) & Rotation of the hand relative to the wrist.  \\
{[} n]domshoulderreloc[xyz] & Inflection (Shoulders) & Translation of the shoulders with respect to the clavicles. \\
torsoreloc[ a][xyz] & Inflection (Torso) & Translation and rotation of the torso trajectory with respect to the hips. \\
headrot[xyz] & Inflection (Head) & Rotation of the head with respect to the neck. \\
\bottomrule
\end{tabular}
    \caption{The 45 MMS columns, grouped by inflection function for brevity, and their description.}
    \label{tab:mms-columns}
\end{table*}

An MMS instance is essentially a table in which each row corresponds to a sign that the avatar is to perform. The first column of the table (\texttt{maingloss}) indicates the gloss-ID \cite{johnston_archive_2010} of a sign to be performed. If limited to this column, the MMS would be equivalent to a gloss sequence.
The remaining 45 columns augment the concept of glosses as explained below.

Table \ref{tab:mms-columns} lists all the columns of an MMS table. They are divided into three groups: Glosses, Timing, and Inflection.

\subsection{Glosses}

In addition to the \texttt{maingloss}, it is possible to specify a \texttt{domgloss} and/or a \texttt{ndomgloss}\footnote{At the moment of writing, this feature is not yet fully integrated in the main branch, but an experimental implementation is present in the \texttt{experiment} folder.}, optionally containing a gloss-ID that overrides the animation of the (non-)dominant arm, from the shoulder to the fingertips\footnote{We are planning to add columns to override (non-)dominant finger configurations, which would enable the expression of classifiers.}. This implements the SL phenomenon where the full body executes a sign while one of the arms might execute in parallel another sign. In fact, when both \texttt{domgloss} and \texttt{ndomgloss} columns are specified, animation data from three signs are played back simultaneously.


Parallel execution of signs is often used in combination with the special gloss-ID \texttt{<HOLD>}, which implements a pause in sign execution. This allows for the blocking of the movement of an arm while the rest of the body keeps signing something else. This is used in SL to keep a reference to spatial locations.

If the \texttt{<HOLD>} gloss-ID is found in the maingloss columns, the avatar "freezes" for the specified amount of time, retaining the position reached by the full body at the end of the previous sign. Together with timing control, this allows for the expression of pauses and rhythm in SL, an element that alone can change the meaning of a sentence (as with the use of commas and periods in written languages).

\subsection{Timing}

\paragraph{Absolute timing}
The sequence of glosses has to be sequentially arranged on a timeline.
The \texttt{framestart} and \texttt{frameend} columns specify the moment on the timeline where the sign starts and ends its execution\footnote{This values used to be specified in \emph{frames}, but the latest implementation interprets it as \emph{seconds}. Hence, these two fields will be renamed in future refactoring.}. Row-by-row, these timestamps are supposed to increase with respect to the previous sign, and the end time being later than the start time; otherwise, results are unpredictable.

Every insulated sign in the dictionary has its ``nominal'' duration (i.e., its recording time). If the difference frameend~-~framestart differs from the original sign duration, the playback will be accelerated or slowed down.

\paragraph{Relative timing}
An alternative way to specify signs timing is through the \texttt{duration} and \texttt{transition} columns. Instead of specifying timestamps on an absolute scale, they allow for express timing in a relative fashion.
The \texttt{duration}, in seconds, is a float number indicating the duration of the sign playback. If the number is followed by a percentage (\texttt{\%}) sign, it indicates a percentage of its nominal playback speed.
The \texttt{transition} column says how many seconds to wait after the end of the execution of the previous sign.
The relative timing method is more human-friendly when composing MMS instances by hand.


%
%
\subsection{Inflections}

Here we list all the inflection functions and the meaning of their parameters.

The function \texttt{(n)domhandreloc} applies a 3D rigid transformation to all points in the \textbf{trajectory of a hand} (dominant or non-dominant) within a sign.
The parameters x/y/z/ax/ay/az/sx/sy/sz specify the translation, rotation, and scale along the three-dimensional axes of the trajectory point with respect to the first point of the trajectory.
With this powerful motion editing, the motion of the hands can be shifted in space (sign relocation), rotated (useful to modify the direction of directional verbs), and scaled (useful to ``shrink and stretch'' a sign).

The function \texttt{(n)domhandrot} applies an additional \textbf{hand rotation} with respect to the wrist. This is useful for changing the orientation of a hand while performing a sign, e.g., to point in a different direction.
The parameters x, y, and z specify the amount of rotation.

The function \texttt{headrot} applies an extra delta angle of \textbf{head rotation} relative to the neck.
The parameters x, y, and z specify additional head yaw, pitch, and tilt.
This is useful, for example, to tilt the head back when performing a question.

The function \texttt{(n)domshoulderreloc} applies a translation to \textbf{shift the shoulders}.
The parameters x, y, and z express the amount of translation.
This is useful, for example, to shrug in case of interrogation or to crouch during role-taking.

The function \texttt{torsoreloc} applies a delta \textbf{translation and rotation to the torso}.
The parameters x/y/z/ax/ay/az specify the amount of translation and rotation.
This motion editing is useful for implementing torso leaning forwards / backwards, and torso orientation to perform role-taking.

\section{Software Architecture}
\label{sec:sw-architecture}

The MMS-Player is implemented as a set of Python functions for the popular Blender\footnote{\url{https://www.blender.org}} 3D authoring tool.
Technically, the MMS-Player is executed by invoking the Blender executable from the command line and specifying the MMS-Player \emph{main.py} entry point as a parameter. This allows for the generation of SL videos without even opening the Blender GUI, which is useful for batch translation processing. An additional \emph{Serve.py} script wraps the Blender execution in a REST API (using the Flask\footnote{\url{https://flask.palletsprojects.com/}} framework) thus allowing for remote online generation services.

\subsection{Main realization cycle}

The main goal of the visualizer, given an input MMS, is to take the sequence of glosses specified in the MMS and compose a timeline with the inflected version of each gloss.
It requires each gloss to be stored in a separate Blender file, and also to have a target character to animate.

The top-level realization process on an MMS instance is composed of the following steps. For each row of the MMS:
    \begin{enumerate}[noitemsep]
        \item Load the gloss animation into an action;
        \item resample the gloss duration according to the timing information;
        \item apply the inflections to the gloss (see later for details);
        \item copy the inflected animation data into the main/final timeline according to the timing information;
        \item finalize the scene (e.g., remove temporary actions, setup lights, set render mode);
        \item if requested, export the final result as FBX, Blender file, MP4 render, animation data.
    \end{enumerate}

\begin{figure}
    \centering
    \includegraphics[width=1.0\columnwidth]{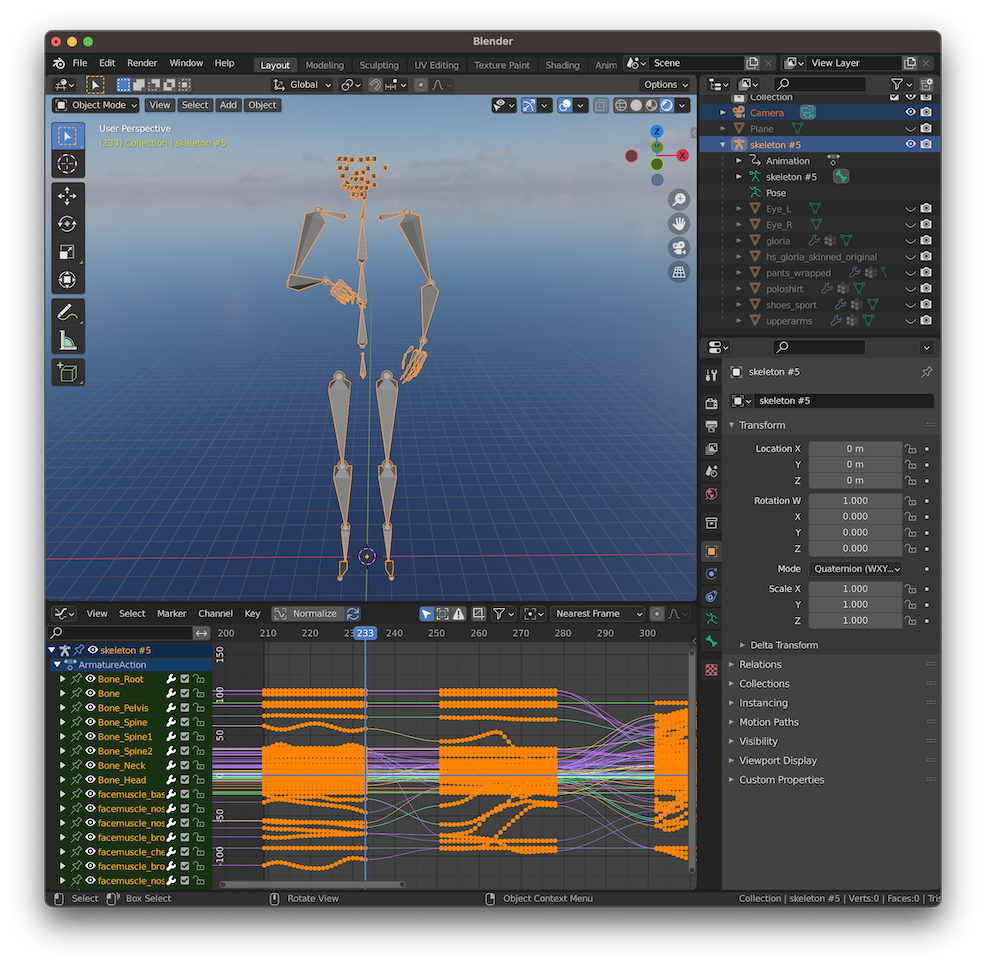}
    \caption{A screenshot of the Blender GUI showing an MMS realization. In this case, only the skeleton of the avatar is shown in order to inspect the motion. At the bottom, in the timeline, it is easy to notice the alternation between high-frame-density gloss animations and procedural transitions.}
    \label{fig:realized-scene}
\end{figure}

Figure \ref{fig:realized-scene} shows a screenshot of the Blender GUI containing a realized MMS instance.

\subsection{Implementation of inflections}

Given the animation of a gloss, an inflection (step 3 of the main cycle) is performed through these steps:

\begin{itemize}[noitemsep]
    \item Create a copy of the original animation curves;
    \item dynamically create an IK controller;
    \item bake the trajectory of the IK target bone into the IK controller;
    \item programmatically edit the motion of the IK controller (e.g., translation offset, rotation offset, ...) using an \texttt{Inflector} (see below);
    \item bake the resulting IK animation back into the armature bone animation curves;
    \item remove the temporary IK controller.
\end{itemize}

When realizing an MMS, this procedure must be applied to each row of the MMS, and for each required inflection.
From a practical point of view, instead of creating and destroying IK controllers at each inflection, the process is optimized by first creating all IK controllers, baking all of them, and then applying the inflections to all of them.
In so doing, only two baking processes (from bones to IK and then back to bones) per row are needed.

\begin{table}
    \centering
    \scriptsize
        \begin{tabular}{l|p{0.5\linewidth}}
        \toprule
        Class & Description \\
        \midrule
\texttt{Inflector(ABC)}              & The top-level abstract interface\\
\texttt{~~GenericInflector}             & Abstract interface implementation, adding fields for all relevant data.\\
\texttt{~~~~LocalRotationTarget}     & The simplest of all inflection strategies, adding only a delta to the local relative rotation. Does not use IK controllers.\\
\texttt{~~~~RelativeLocRotTarget}    & Applies a delta translation and rotation to an IK controller. The delta values are relative to a given root, which is also the root of IK chain.\\
\texttt{~~~~RelativeRotTarget}           & Applies a delta rotation to an IK controller. The delta value is relative to a given root, which is also the root of the the IK chain.\\
\texttt{~~~~TrajectoryTarget}        & The most complex of all inflections. Applies a full 3D rigid transformation (translation, rotation, non-uniform scaling) to all points of the trajectory of an IK controller.\\
    \bottomrule
    \end{tabular}
    \caption{The hierarchy of inflection classes.}
    \label{tab:inflection-classes}
\end{table}

Table \ref{tab:inflection-classes} shows the hierarchy of inflection classes.
The core of the inflection is performed by subclasses of the \texttt{Inflector} abstract class, whose hierarchy of inflection sub classes is shown in table \ref{tab:inflection-classes}.
An Inflector subclass knows how to inflect something by providing a concrete implementation of the \texttt{.inflect()} method.
We refer to concrete subclasses as ``inflection strategies'', i.e., how to perform an inflection.

\begin{table*}
\small
\centering
\begin{tabular}{l|l|l|l}
\toprule
MMS column & Inflection class & Controlled bone(s) & Relative bone \\
\midrule
{[} n]domhandreloc[ as][xyz] & TrajectoryTarget & Bone\_(L)R\_Hand & Bone\_Spine2 \\
{[} n]domhandrot[xyz] & LocalRotationTarget & Bone\_(L)R\_Hand & Bone\_Spine2 \\
{[} n]domshoulderreloc[xyz] & RelativeLocTarget & Bone\_(L)R\_Clavicle & Bone\_Spine2 \\
torsoreloc[ a][xyz] & RelativeLocRotTarget & Bone\_Spine2 & Bone\_Pelvis \\
headrot[xyz] & RelativeRotTarget & Bone\_Head & Bone\_Spine2 \\
\bottomrule
\end{tabular}
    \caption{The correspondence between MMS inflections and the  strategies (w/ parameters) to implement them.}
    \label{tab:mms-inflection-columns}
\end{table*}

An inflection strategy (i.e., a concrete Inflector) is a generic way of inflecting, and can be reused for several body parts and for several sign language inflections, but it is conceptually detached from a specific application domain (e.g., a \texttt{TrajectoryTarget} could be used to amplify the motion of the tail of an animal).
The mapping between a specific MMS inflection and the inflection strategy is listed in table~\ref{tab:mms-inflection-columns}.

It is worth noting that the root of most IK controllers is set to the torso \texttt{Bone\_Spine2}, while only the inflection of the torso is relative to the hips (\texttt{Bone\_Pelvis}).
This means that, for example, when inflecting the trajectory of the hands to move a sign towards the right, even if the torso is rotated towards the left, the sign will be still relocated towards the right shoulder of the avatar, regardless of the amount of torso rotation. As if the torso is always ``carrying'' the signing space in front of it.
This choice is the result of a set of tests and trials conducted during the development phase.

\section{Preliminary Evaluation}
\label{sec:evaluation}

Our work also features a preliminary user study, the purpose of which was to gather initial feedback from a selected group of experts. This study was crucial in identifying challenges and assessing the current state of the project.

\begin{figure}
    \centering
    \includegraphics[width=1.0\linewidth]{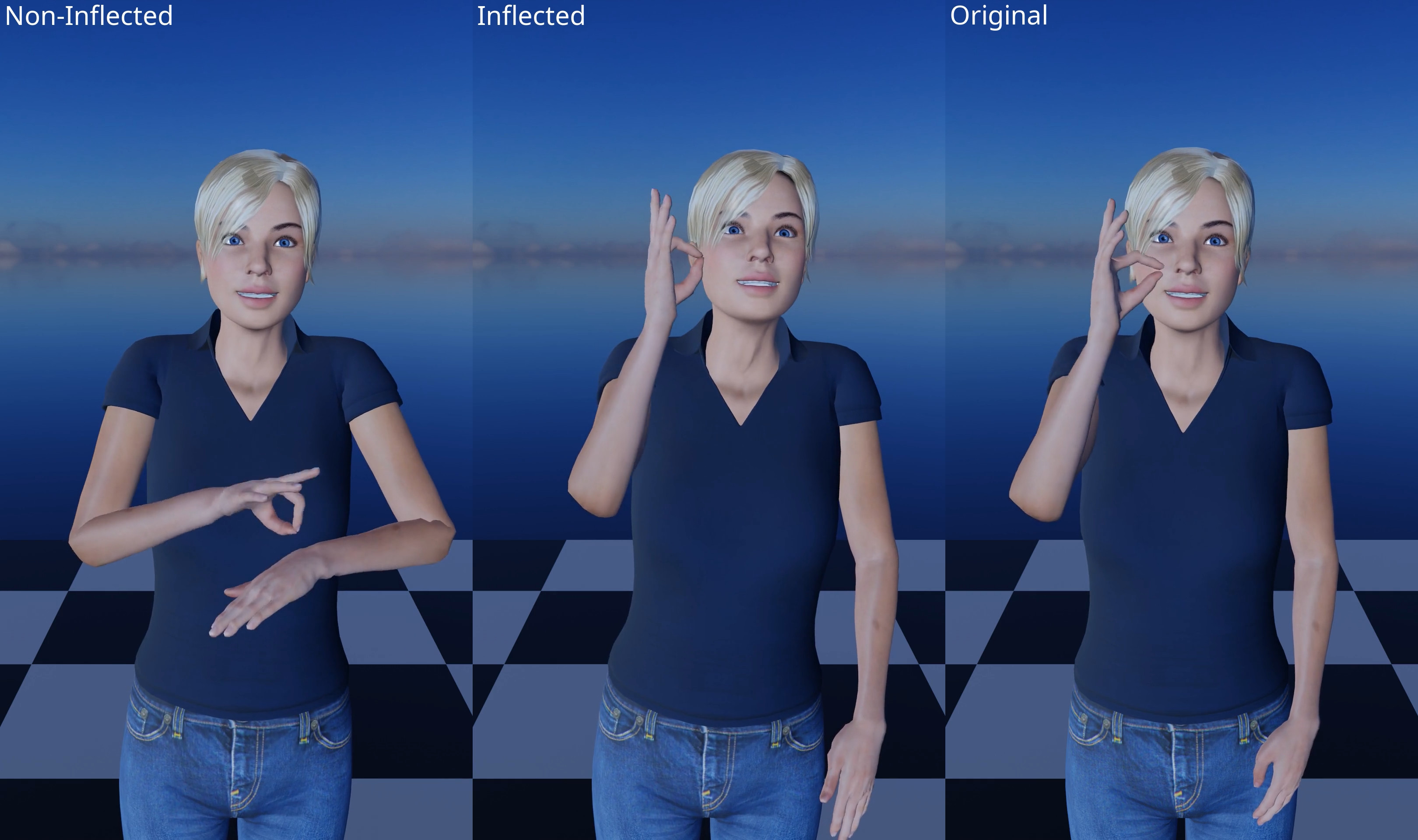}
    \caption{A frame from one of the evaluation videos.}
    \label{fig:evaluation-screenshot}
\end{figure}

As study material, we prepared three different SL video sentences, containing 55 unique glosses in total.
Each video showed an avatar signing under three conditions: Non-Inflected on the left, Inflected in the middle, and Original on the right. See an example in figure \ref{fig:evaluation-screenshot}.

The Original condition is the animation of the avatar using the motion capture data of the full sentence.
The Non-Inflected animation is the concatenation of the glosses of the sentences, only modulated in speed to match the duration of the glosses of the original sentence, but without any other inflection.
The Inflected condition applies the inflection parameters to each gloss of the sentence to match as closely as possible the body motion of the original sentence. The optimal inflection parameters were automatically computed using the method described by Mishra~\cite{mishra_inflection_2023}.

As can be seen in figure \ref{fig:evaluation-screenshot}, during the execution of gloss ``minutes'', the original insulated sign is heavily inflected to match more closely the specific variation used by the interpreter in the sentence.

We prepared a questionnaire to answer the following research question:
applying inflections improves or worsens the quality of animation?

For each video, users were asked to answer two questions:
\begin{enumerate}
        \item How would you rate the Naturaleness, Grammatical Correctness, and Understandability of the avatar animation for the left, center, and right videos?
        \item Could you please explain the reason behind your rating?
\end{enumerate}

While the last answer was given as free text, the first questions were evaluated with a 7-point Likert-like scale in terms of Naturalness, Grammatical Correctness, and Understandability, using the following terms.

\begin{enumerate}[noitemsep]
    \item Naturalness of the avatar animation: Entirely Unnatural, Largely Unnatural, Moderately Unnatural, Ambiguous, Moderately Natural, Largely Natural, Entirely Natural.
    \item Grammatical correctness of the avatar animation: Completely Erroneous, Significantly Erroneous, Mildly Erroneous, Ambiguous, Mildly Accurate, Significantly Accurate, Completely Accurate.
    \item Understandability of the avatar animation: Wholly Unintelligible, Predominantly Unintelligible, Partially Unintelligible, Ambiguous, Partially Clear, Predominantly Clear, Wholly Clear.
\end{enumerate}

\begin{figure}
  \centering
  \includegraphics[width=\linewidth]{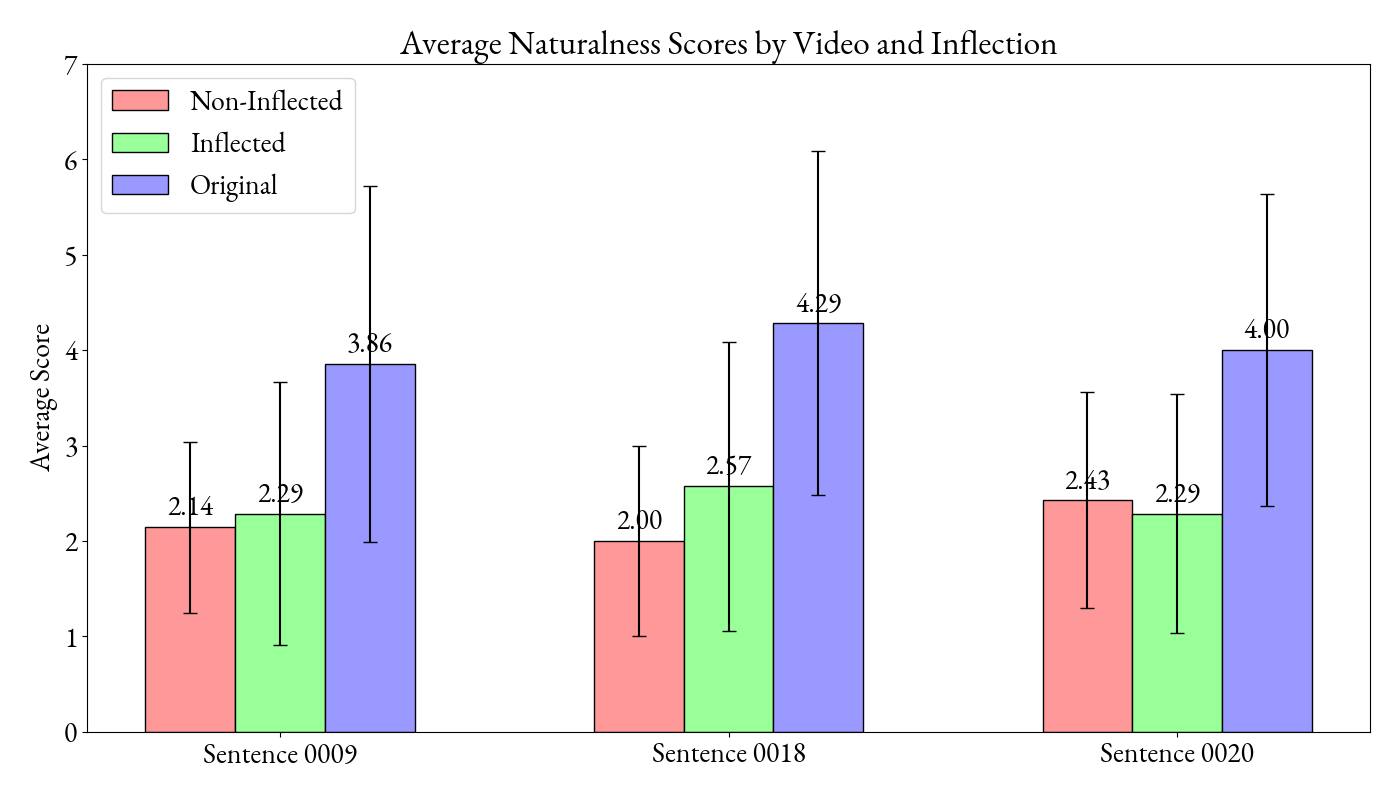}
  \includegraphics[width=\linewidth]{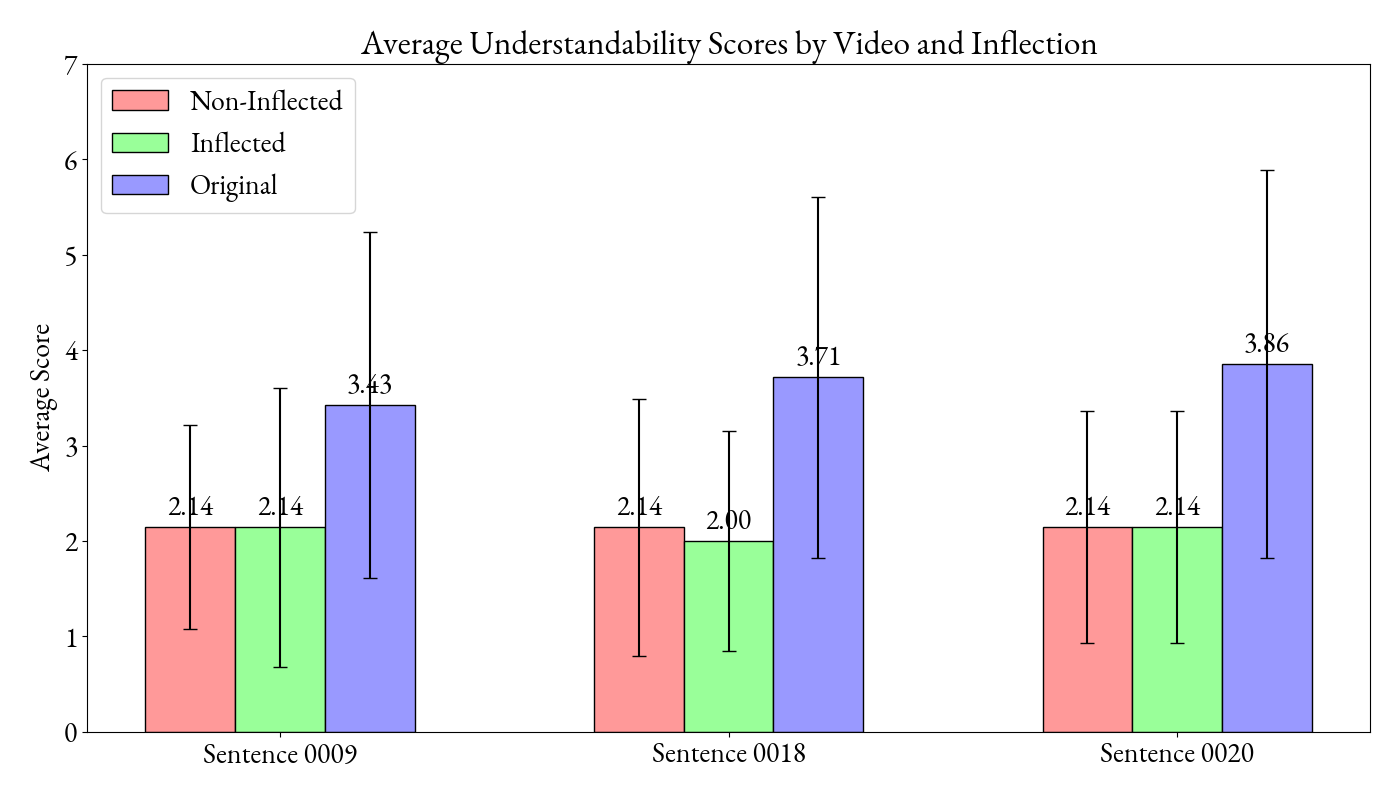}
  \includegraphics[width=\linewidth]{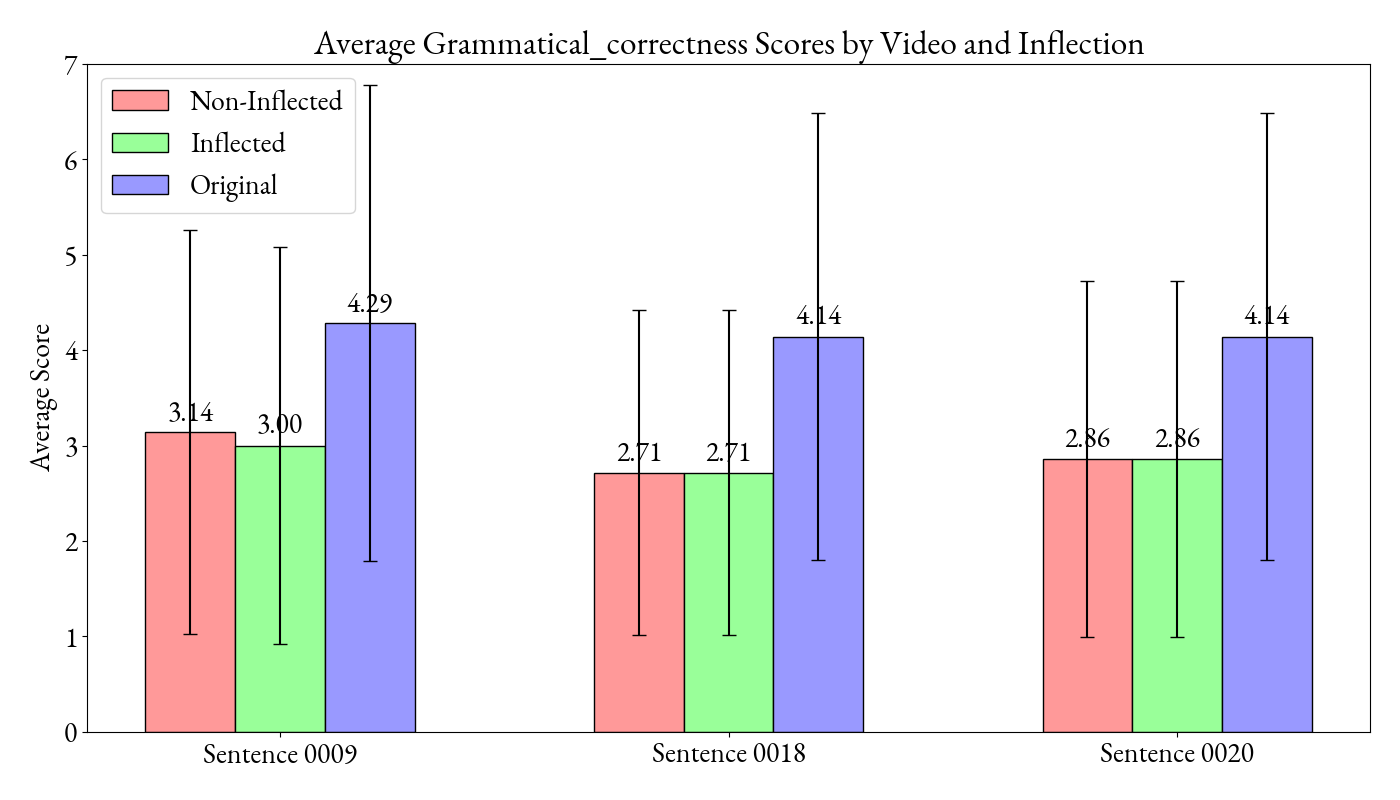}
  \caption{Results for users evaluation on Naturalness, Understandability, Grammatical correctness.}
  \label{fig:user-study-results}
\end{figure}

The results of the study, based on votes collected from 5 subjects, are plotted in figure \ref{fig:user-study-results}.

Without the need to run statistical tests, it is clear that the votes for the Inflected condition are not better than the version with the Non-Inflected condition.
However, it is also noticeable that the baseline condition Original received scores very close to the neutral 4.0 score, but with a high variance.
An analysis of free comments helped us to better understand users' feedback and identify the most important elements to improve the system.

One of the most notable comments came from a native DGS subject. Despite being clearly asked to ignore the face because the prototype worked only on the body, the native subject commented that ``I cannot judge because the face is not animated'' and gave consistently low scores.

For \textbf{Naturalness}, users commented that even the original motion capture video does not look natural, because of the abrupt motion of the torso leaning in the Inflected videos, and because in the Non-Inflected video the torso appeared too rigid.
This is due to an unfortunate choice made at the beginning of the project: to skip the animation of the hips. When animating sign language, it seemed obvious to skip the animation of the lower part of the body and focus on the animation of the upper body. However, in reality, when the torso is leaning the hips are normally performing a counter-rotation which compensates partially the rotation of the lower part of the spine and keeps the torso at a more natural inclination. By removing the hip rotation, the leaning of the torso is exaggerated.

Also, users criticized the timing and signing speed. This is a consequence of a choice that aimed to reduce the number of videos shown to subjects. We decided to compose one video with all three conditions at once. To normalize the duration of the signing, we used the timing of the Original video for the two other conditions. The result is that all of the vocabulary signs are reproduced at a speed that is roughly the double of their nominal speed, resulting in a body dynamics that goes beyond credibility.

Concerning \textbf{Understandability}, most users reported ``disruption of flow'' as a reason for not being able to comprehend the sentence. By ``flow'' is meant the jittering effect when procedurally generating the animation to transition from the end of a sign to the beginning of the next sign. Hence, it seems that the flow of body motion is rather important for the understanding of the sentence, and not only the correct playback of the single signs.

Finally, regarding \textbf{Grammatical Correctness} users reported that the concept of grammar crucially depends on the inflection of facial expressions.

\section{Conclusions and future work}
\label{sec:conclusions}

We presented the motivation and the architectural elements of the implementation of the MMS-Player: an open source GPL-3.0-licensed set of Python scripts for the Blender 3D editor that allows one to generate sign language animations. The MMS-Player takes into account the fact that sign language cannot be fully synthesized with a mere concatenation of motion-captured insulated signs. It proposes a solution that allows for the realization of all known SL phenomena from an animation point of view, with a level of abstraction that lies below grammar or phonetic formalizations.\\

The MMS-Player is currently in a development stage, and a preliminary user study allowed us to identify the most critical points that need to be addressed for future software development as well as for the design of future experiments.

First, facial animation is a fundamental and integral part of sign language communication: without implemented facial animation, it does not even make sense to conduct more user studies. 

Second, when recorded in isolation, signs cannot be played back at the speed naturally used in fluent signing. To what magnitude sign playback can be accelerated needs to be investigated.

Third, hip motion must be retained from the motion capture data and supported by the system, possibly also supported by an additional inflection function to control translation and rotation.

Finally, transitions between signs must be handled by ad hoc routines. The default animation interpolation technique of Blender is not sufficient to realize believable animations, which are currently ``disrupting the flow'' of sentences.

Our plan is to add to the MMS additional information about expressed  emotions \cite{nunnari_emotions_2024} and 
address the above-mentioned issues to further develop a product that is better accepted, and hopefully adopted, by the research and the deaf communities.

%
%
%
\section*{Acknowledgments}
This research was funded by the German Ministry of Education and Research (BMBF) through the AVASAG project (grant number 16SV8491) \cite{nunnari2021AT4SSL-AVASAG,bernhard22PETRA-AVASAG} and through the BIGEKO project (grant number 16SV9093).
Thanks to Charamel GmbH (\url{https://charamel.com}) for allowing the distribution of the avatar asset in the open source repository.

\bibliographystyle{plain}
\bibliography{nunnari,SLTAT2025-MMSPlayer}

\end{document}